\documentclass[useAMS,usegraphicx, manuscript, usenatbib]{mn2e}
\usepackage{aas_macros}

\usepackage{amsmath}
\usepackage{amssymb}
\usepackage[utf8x]{inputenc}
\usepackage{graphicx}
\usepackage{epsfig}
\usepackage{latexsym}
\usepackage{color}
\usepackage{dcolumn}% Align table columns on decimal point
\usepackage{bm}% bold math
\usepackage{natbib}
\usepackage{setspace}
\usepackage{subfigure}
 \usepackage{psfrag}

\begin{document}
%\title{The influence of dark matter halo properties on primordial star formation} 
\title{Dark Matter Halo Environment for Primordial Star Formation} 

\author[R. S. de Souza,  B. Ciardi, U. Maio, A. Ferrara]
{R. S. de Souza  $^{1,2,3}$, B. Ciardi$^{2}$,  U. Maio$^{4}$, A. Ferrara$^{5}$
\\
$^{1}$Korea Astronomy \& Space Science Institute, Daejeon 305-348, Korea\\
$^{2}$Max-Planck-Institut f\"ur Astrophysik, Karl-Schwarzschild-Str. 1, D-85748 Garching, Germany\\
$^{3}$IAG, Universidade de S\~ao Paulo, Rua do Mat\~ao 1226, Cidade Universit\'aria,
CEP 05508-900, S\~ao Paulo, SP, Brazil\\
$^{4}$Max-Planck-Institut f\"ur extraterrestrische Physik,
Giessenbachstra{\ss}e 1, D-85748 Garching bei M\"unchen, Germany\\
$^{5}$Scuola Normale Superiore, Piazza dei Cavalieri 7, 56126 Pisa, Italy
}

% \date{Accepted -- Received  --; in original form 2010 February 1}
% \date{Accepted -- Received  --}

\pagerange{\pageref{firstpage}--\pageref{lastpage}} \pubyear{2010}

\maketitle
\label{firstpage}

\begin{abstract}
We study  the statistical properties (such as shape and spin) of high-\emph{z} halos likely hosting the  first (PopIII) stars
with cosmological simulations including detailed gas physics. In the redshift range considered ($11 < z < 16$) the average 
sphericity  is $\langle s \rangle = 0.3 \pm 0.1$, and for more than 90\% of halos  the triaxiality parameter is $T \lesssim 0.4$, 
showing a clear preference for oblateness over prolateness. Larger halos in the simulation tend to be both more spherical and prolate: 
we find $s \propto M_h^{\alpha_s}$ and  $T \propto M_h^{\alpha_T}$, with $\alpha_s \approx 0.128$ and  $\alpha_T= 0.276$ 
at \emph{z} = 11.  The spin distributions of dark matter and gas are considerably different at $z=16$, with the baryons rotating 
slower than the dark matter. At lower redshift, instead, the spin distributions of dark matter and gas track each other almost 
perfectly, as a consequence of a longer time interval available for momentum redistribution between the two components.
The spin of both the gas and dark matter follows a lognormal distribution, with a mean 
value at $z=16$ of $\langle \lambda\rangle =0.0184$, virtually independent  of halo mass. This is in good agreement with 
previous studies. Using the results of two feedback models (MT1 and MT2) by McKee \& Tan (2008) and mapping our halo 
spin distribution into a PopIII IMF, we find that at high-$z$ the IMF closely tracks the spin lognormal distribution. 
Depending on the feedback model, though, the distribution can be centered at  $\approx 65 M_\odot$ (MT1) or $\approx 140 
M_\odot$ (MT2).  At later times, model MT1 evolves into a bimodal distribution with a second prominent peak located 
at $35-40 M_\odot$ as a result of the non-linear relation between rotation and halo mass. We conclude that the dark matter 
halo properties might be a key factor shaping the IMF of the first stars.  
\end{abstract}

%\begin{keywords}
%cosmology: early Universe, large-scale structure of Universe; methods: N-body simulations 
%\end{keywords}

\section{Introduction}

The  formation of first, metal-free (often referred to as PopIII)  stars in the Universe represents a milestone during cosmic 
evolution, marking the end of the Dark Ages and producing the first heavy elements \citep{ciardi2005,Yoshida2008,bromm09,Bromm2011,desouza2011a,rafael2012,Johnson2012}.  
Thus, a key problem in physical cosmology is to understand the origin and evolution of such objects, born out of the
pristine conditions leftover by the Big Bang. More specifically, the most urgent question concerns their Initial Mass 
Function (IMF), which, despite its relevance, remains at best a poorly known quantity due to the lack 
of  direct observations. Consequently, our knowledge is based mainly on theoretical models (e.g., \citealt{Dopcke2012}). 

Until recently, studies based on the standard $\Lambda$CDM cosmological model\footnote{Throughout the paper we assume a standard $\Lambda \rm CDM$ cosmological model, with
current total-matter density parameter $\Omega_{m}=0.3$, cosmological constant density parameter 
$\Omega_{\Lambda}=0.7$, baryonic-matter density parameter $\Omega_{b}=0.04$, expansion rate 
in units of 100~km~s$^{-1}$~Mpc$^{-1}$, $h = 0.7$, spectral normalization $\sigma_8=0.9$, and 
primordial spectral index $n=1$.} for structure formation
predicted that the first stars formed when the age of the Universe was less than a few hundred million years,
and that they were predominantly massive \citep{abel2002,omukai2003,yoshida2006}.
\citet{clark2011}, \citet{greif2011} and \citet{Prieto2011} have now performed cosmological simulations using a 
sink particle technique to follow the evolution of a primordial protostellar accretion disk. They find that instead of 
forming a single massive object, the gas typically fragments  into a number of protostars with a range of different masses. 
However, high resolution radiation-hydrodynamics simulations by \citet{hosokawa2011} indicate
a typical mass of PopIII stars of $\sim 43 M_{\odot}$.  Even more recently, similar results was found  by \citet{Stacy2012a}, which  suggests radiative feedback will lower the final mass attainable by a PopIII star. With a estimative of $30 M_{\odot}$ for a PopIII mass as a lower limit. \citet{Greif2012},  using  a different numerical
scheme without  need to insert sink particles, found that the keplerian disc around the primary protostar fragments into a number of 
secondary protostars, confirming previous results using sink particles. Clearly, these theoretical results are far  from  
being conclusive, mostly due to the astonishing difficulty involved in simulating the large dynamical range required and
the complex physics involved. Note that these studies, with few exceptions  (e.g. \citealt{Turk2012}),  largely neglect the 
possible effects of a magnetic field in the fragmentation properties of the gas.   

In spite of this unsettled situation, a broad consensus exists on the fact that rotation of the protogalactic
cloud is the key factor in determining the final outcome of the collapse. The importance of rotation has been
fully appreciated after the fundamental paper by \citet[][MT08]{mckee2008}, who studied the dependence of  
primordial protostar accretion in the presence of rotation (fostering the formation of an accretion disk) and radiative 
feedback from the protostar. In their study MT08 concluded that the  final stellar mass depends mainly on the 
entropy of  the gas accreting from large radii, as well as its specific angular momentum. As the gas is bound
in the gravitational potential of the dark matter halo, it follows that the angular momentum of the gas is linked to
that of the parent halo. Hence, it is important to turn our attention to the properties of the dark matter halos
that host the first stars to put the entire problem on a solid basis.

The distribution of the angular momentum enters a broad range of problems, from the halo mass function itself 
(e.g. \citealt{Maccio2007,DelPopolo2009}), to the formation and evolution of central black holes 
(e.g. \citealt{Eisenstein1995,Volonteri2005,Volonteri2010}), to semi-analytical models of galaxy formation 
(e.g. \citealt{Benson2012}). Simulations of  isolated dwarf galaxies suggest  that angular momentum  leads to more continuous star formation histories than non-rotating cases by preventing large-scale oscillations in the star formation rate \citep{Schroyen2011}.
 For this reason, much effort has been done to explore the spin 
distribution of halos in different redshift and mass ranges, to characterize its probability distribution, 
as well as the dependence on mass, shape, merger rate  and other halo  properties  
(e.g. \citealt{Bett2007,DOnghia2007,Antonuccio2010}).   

Another fundamental characteristic of halos is their shape.  Spherical halos are very rare, 
and their collective properties cannot be approximated using spherical symmetry \citep{Allgood2006}. They are
usually described in terms of ellipsoids characterized by three principal axis. Many authors 
have explored ways to estimate the shape of halos and the correlations with other halo 
properties (e.g. \citealt{Zemp2011}).  
Using a pure dark matter N-body simulations, \citet{Jang-Condell2001}  analyzed the primordial halo characteristics  and found no significant difference between their results and simulations of large-scale structure formation at low redshift.   
 \citet{Kazantzidis2004} found that halos in cosmological simulations including 
gas cooling are considerably more spherical than those found in adiabatic simulations.  This shows that the inclusion 
of detailed gas physics is fundamental, since the back-reaction effects of baryons on dark matter halos change their density profiles as well as their mass distribution \citep{Cui2012}.  Observationally, weak gravitational 
lensing is probably the most common approach to reconstruct the shape of halos 
(e.g. \citealt{Corless2008,Bett2012,vanUitert2012}). The method  does not depend on the presence of optical 
tracers and can be applied  to a large range of scales.

The purpose of the present paper is to explore  the properties (such as spin and shape) of the high-$z$ halos likely to host the first stars, with an unprecedented inclusion of detailed gas
physics. Although previous works that have investigated 
the correlation between halo parameters exist (e.g., \citealt{Jeeson2011,Skibba2011}),  this is the first attempt to 
determine the characteristics and correlations of the low-mass end of the halo mass function including gas physics.

As a final step, we embed our results into theoretical models of first star formation to unveil the links between 
dark matter halo properties and the PopIII IMF.

The outline of this paper is as follows. In Sec.  \ref{sec_sim}, we briefly describe the N-body/hydrodynamical 
simulations used to derive the halo properties;  Sec. \ref{sec_properties} describes the methodology used to calculate 
the quantities of interest (halo spin and shape). In Sec. \ref{sec_results}, we show the results and provide useful fits 
for correlations between different quantities. Finally, Sec. \ref{sec_IMF} discusses the possible implications for 
the PopIII IMF. Sec. \ref{sec_disc} contains a summary of the results. 

%%%%%%%%%%%%%%%%%%%%%%%%%%%%%%%%%%%%%%%%
\section{Simulations}
\label{sec_sim}
%%%%%%%%%%%%%%%%%%%%%%%%%%%%%%%%%%%%%%%%

We analyze the output of the  N-body/hydrodynamical simulations described in  \citet{maio2010},
which were performed using the Gadget-2 code \citep{Springel2005}. The simulations include the 
evolution of e$^-$, H, H$^+$, H$^-$, He, He$^{+}$, He$^{++}$, H$_2$, H$_2^+$, D, D$^+$, HD, HeH$^+$  
\cite[][]{yoshida2003,maio2006,maio2007,maio2009}, PopIII and PopII/I star formation and metal 
pollution \cite[][]{tornatore2007}, gas cooling from resonant and fine-structure lines 
\cite[][]{maio2007,maio2009} and feedback effects \cite[][]{springel2003}. 
 There are evidences for the existence of a critical metallicity, $Z_{crit}$, which allow the transition between PopII/I and Pop III star formation modes \citep{Omukai2000,Bromm2001}. The transition from the PopIII to the PopII/I regime is determined by the  value of the gas 
metallicity, $Z$, compared to the critical value $Z_{crit}=10^{-4}Z_\odot$.  The value of this minimum level is very uncertain, but is likely to be between $10^{-6}$ and $10^{-4} M_{\odot}$ and can be strongly dependent  of  the efficiency of dust formation in first-generation supernova ejecta \citep{Schneider2003,Schneider2006} and the  fine-structure line cooling of singly-ionized carbon or neutral atomic oxygen \citep{bromm2003}.  
If $Z<Z_{crit}$ a Salpeter IMF (i.e. with a slope of -1.35) is assumed in the mass range 100-500~$M_\odot$, 
otherwise  the same Salpeter slope is adopted in the mass range 0.1-100~$M_\odot$.
The chemical model follows the detailed stellar evolution of each SPH particle. At every time-step, 
the abundances of various heavy elements (C, O, Mg, S, Si and Fe) are consistently derived, according to the 
lifetimes and metal yields of the stars in the given mass range.

The cosmological field is sampled at redshift $z=100$, with dark-matter and baryonic-matter species.
We consider  snapshots in the   range $11 < z < 16$, within a cubic volume of comoving side  
1 Mpc $h^{-1}$ and   320$^{3}$ particles per gas and dark matter species, corresponding to particle masses 
of 116 $M_{\odot} h^{-1}$ and 755 $M_{\odot} h^{-1}$, respectively.
The identification of the simulated objects is done by applying a Friends of Friends (FoF) technique; 
substructures are identified by using a SubFind algorithm \citep{Dolag2009}, which discriminates among 
bound and non-bound particles. For more details on the simulations we refer the reader to the original 
paper \citep{maio2010}.

%%%%%%%%%%%%%%%%%%%%%%%%%%%%%%%%%%%%%%%%%%%
\section{Halo properties derivation}
\label{sec_properties}
%%%%%%%%%%%%%%%%%%%%%%%%%%%%%%%%%%%%%%%%%%%

In the following we describe the method used to derive the halo properties of interest here, as its shape and 
spin, along with a number of ancillary quantities defined below.

%%%%%%%%%%%%%%%%%%%
\subsection{Shape}
%%%%%%%%%%%%%%%%%%%

The halo shape is estimated based on its mass distribution, which can be directly derived 
using the eigenvalues of the  inertia tensor $\mathcal{I}$ (e.g., \citealt{Springel2004,Allgood2006,Bett2007}),
\begin{equation}
\mathcal{I}_{jk}= \sum_{i=1}^{N}m_i(\mathbf{r}_{i}^2\delta_{j,k}-r_{i,j}r_{i,k}),  
\end{equation}
where $\mathbf{r_i}$  and $m_i$ are  the position vector and mass of the $i$-th particle, $\delta_{j,k}$ 
is the Kronecker  delta and the sum is performed over the total number of particles inside  the halo, $N$. 

An alternative way to measure the shape is by using the second moment of the mass distribution, 
i.e. the shape tensor
\begin{equation}
\mathcal{S}_{jk}= \frac{1}{N}\sum_{i=1}^{N}r_{i,j}r_{i,k}.  
\label{eq:shape}
\end{equation}
As the eigenvalues of $\mathcal{S}$ and $\mathcal{I}$ are the same, the two definitions are
totally equivalent for  halo shape studies. We then restrict our discussion to the shape tensor.

The eigenvalues of the diagonalised shape tensor define an ellipsoid, which represents the equivalent 
homogeneous shape of the halo in terms of  the principal axis ratios, with the convention 
$a \geqslant b \geqslant c$.
It is customary to refer to the axis ratios  in terms of sphericity, $s = c/a$ (with $s=0$ meaning aspherical 
and $s=1$ spherical), oblateness, $q = b/a$ and prolateness, $p= c/b$. With these definitions, 
the  triaxiality parameter can be conveniently written as
\begin{equation}
T = \frac{1-q^2}{1-s^2}; 
\end{equation}
hence a prolate (oblate) halo has $T = 1$ ($ T= 0$).  

A slightly different way to calculate the shape tensor was introduced by \citet{Allgood2006}, 
\begin{equation}
\mathcal{S}_{jk}= \frac{1}{N}\sum_{i=1}^{N}\frac{r_{j,i}r_{k,i}}{d_i^2}, 
\end{equation}
where $d_i^2 = x_i^2+y_i^2/q^2+z_i^2/s^2$ is the elliptical distance in the eigenvector coordinate 
system from the center to the $i$-th particle and $q$ and $s$ are found iteratively. More generally, 
a weight factor, $w(r)$, can be introduced 
\begin{equation}
\mathcal{S}_{jk}= \frac{1}{N}\sum_{i=1}^{N}\frac{w(r)r_{j,i}r_{k,i}}{d_i^2}.
\end{equation}
A critical review of the different approaches can be found in \citet{Zemp2011}. They explore six 
methods, differing by the integration volume and the choice of the weight functions,  
$w(r) = 1, w(r) = 1/r^2, w(r) = 1/d_i^2$. They conclude  that using weights can introduce a systematic bias 
in the measured axis ratios, in addition to blurring the physical interpretation of the shape tensor. 
Thus, using $w(r) = 1$  and integrating over the enclosed ellipsoidal volume appears to be the most 
unbiased method. This choice is also  preferred for halos with lower number of particles and if  the main 
interest is in deriving the global rather than local (i.e. as  a function of distance) shape. Hereafter, all calculations are 
done using eq. \ref{eq:shape} integrated over the enclosed ellipsoidal volume. 

%%%%%%%%%%%%%%%%%%%% FIGURE 1 :: START %%%%%%%%%%%%%%%%%%%%%%%%%%%%%
\label{sec_data}
\label{sec_res}
\begin{figure}
\psfrag{xlabel}[c][c]{$\log{M_{h}} ~(M_{\odot})$}
\psfrag{z = 11}[c][c]{$ z =11 ~$}
   \psfrag{z = 16}[c][c]{$ z =16 ~$}
    \psfrag{s=c/a}[c][c]{$ s$}
    \psfrag{q=b/a}[c][c]{$q=b/a$}
     \psfrag{p=c/b}[c][c]{$p=c/b$}
     \psfrag{T}[c][c]{${\rm T}$}
     \psfrag{4.5}[c][c]{$4.5$}
     \psfrag{5}[c][c]{$5.0$}
     \psfrag{6}[c][c]{$6.0$}
     \psfrag{7}[c][c]{$7.0$}
     \psfrag{8}[c][c]{$8.0$}
      \psfrag{5.0}[c][c]{$5.0$}
      \psfrag{5.5}[c][c]{$5.5$}
     \psfrag{6.0}[c][c]{$6.0$}
     \psfrag{6.5}[c][c]{$6.5$}
     \psfrag{7.0}[c][c]{$7.0$}
     \psfrag{7.5}[c][c]{$7.5$}
         \psfrag{0.0}[c][c]{$0$}
            \psfrag{0.1}[c][c]{$0.1$}
               \psfrag{0.2}[c][c]{$  $}
      \psfrag{0.3}[c][c]{$0.3 $}
     \psfrag{0.4}[c][c]{$$}
        \psfrag{0.5}[c][c]{$ 0.5$}
     \psfrag{0.6}[c][c]{$  $ }
        \psfrag{0.7}[c][c]{$0.7$}
     \psfrag{0.8}[c][c]{$ $}

\centering
%\begin{tabular}{cc}

    \includegraphics[trim = 5mm 0mm 4mm 18mm, clip, width=1.1\columnwidth]{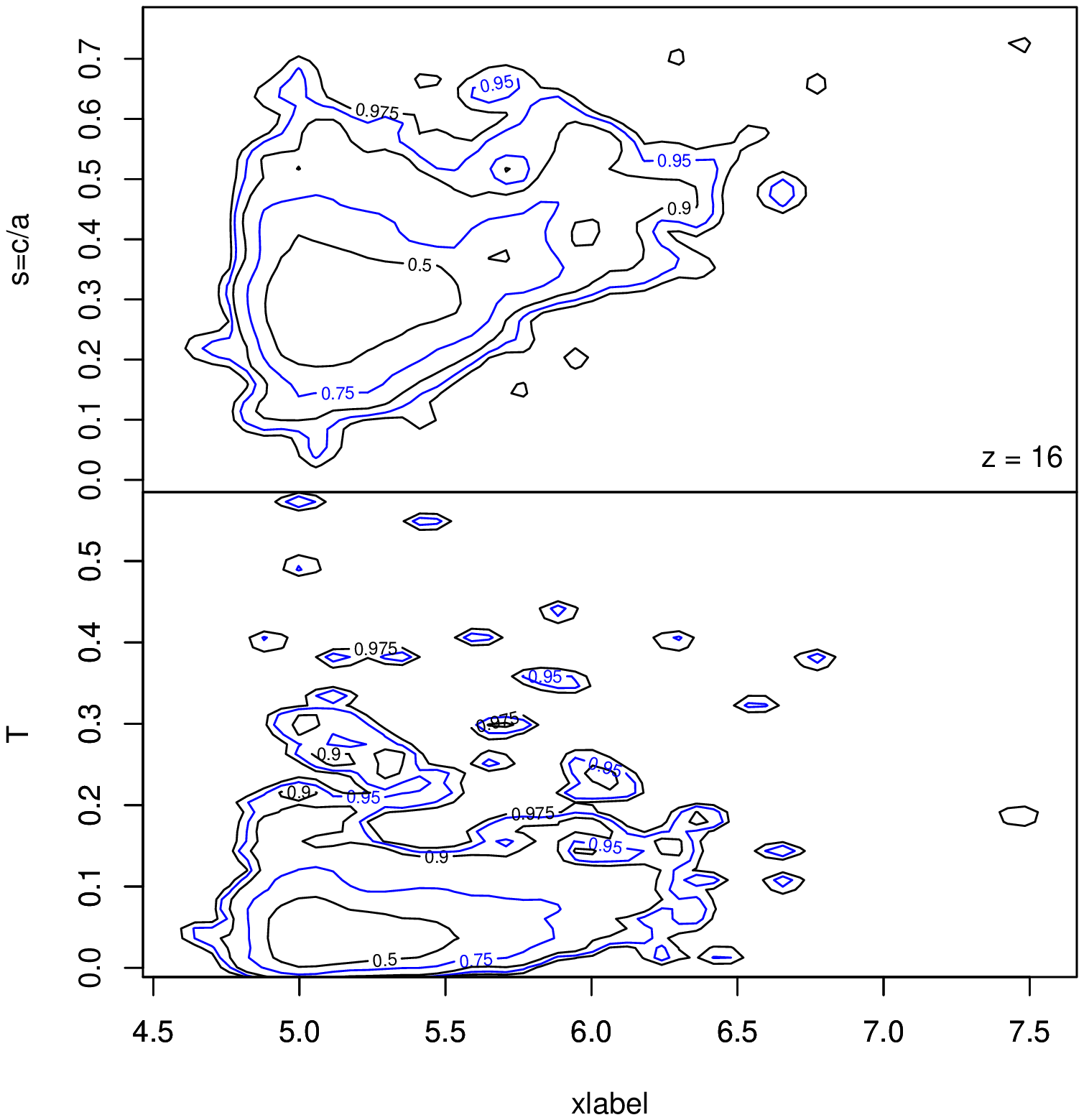}\\ 
     \includegraphics[trim = 5mm 0mm 4mm 10mm, clip,width=1.1\columnwidth]{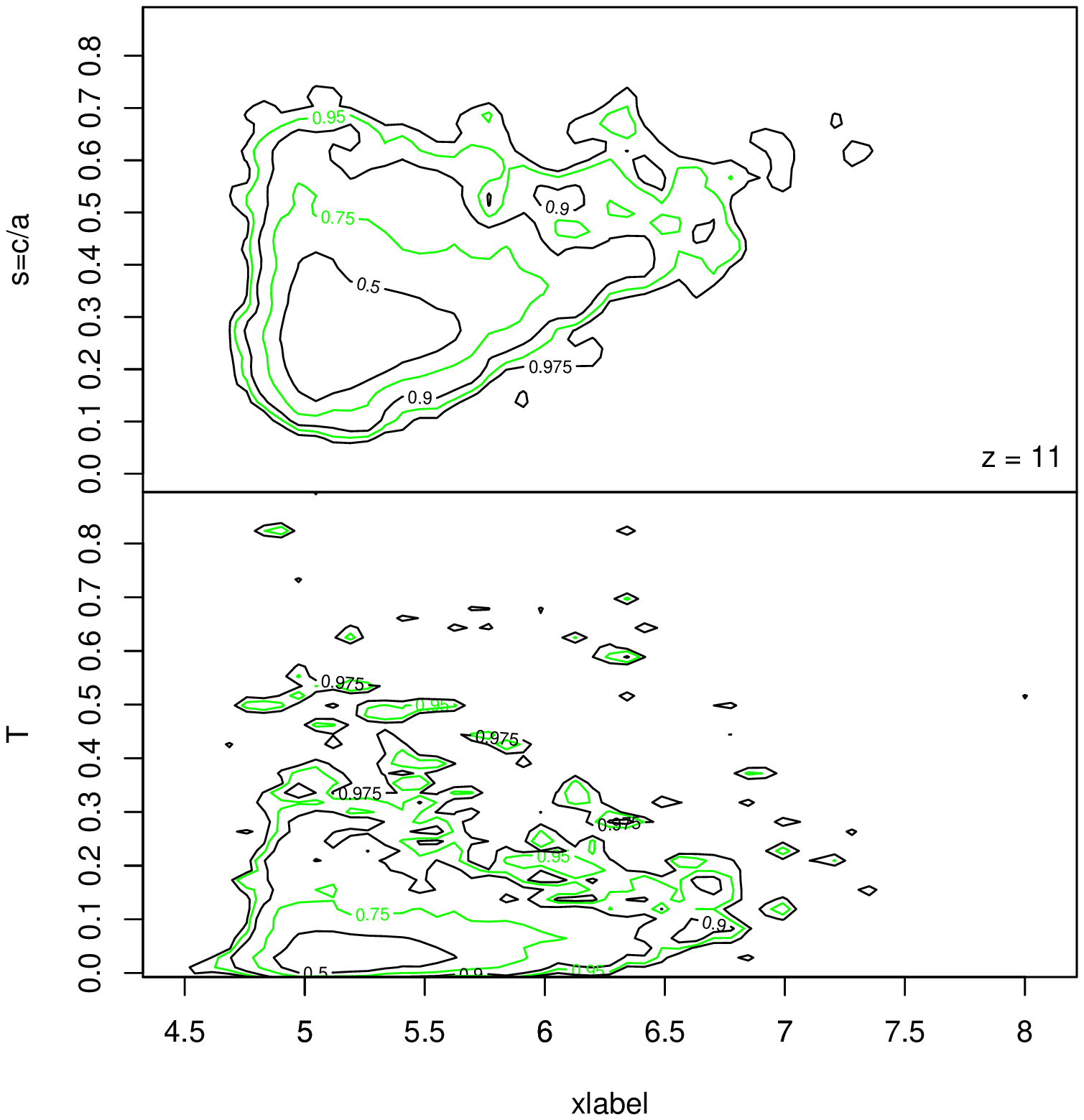}
   % \end{tabular}
   \caption{Sphericity $s$ (upper panels), and triaxiality $T$ (lower panels),
      as a function of total halo mass for halos at $z=16$ (upper figure) and $z=11$ (lower). The contour 
      levels represent 50\%, 75\%, 90\%, 95\% and 97.5\% of the sample.
      }
\label{fig:shape16}
\end{figure}
%%%%%%%%%%%%%%%%%%%% FIGURE 1 :: END %%%%%%%%%%%%%%%%%%%%%%%%%%%%%

%%%%%%%%%%%%%%%%%%%%%%%
\subsection{Spin}
%%%%%%%%%%%%%%%%%%%%%%%

The spin parameter is a measure of the amount of coherent rotation in a system compared to random motions. 
For a spherical object, it corresponds approximately to the ratio of its own angular velocity to the angular velocity 
needed for it to be supported against gravity solely by rotation (see e.g. Padmanabhan 1993). The halo spin can 
be characterized by a dimensionless parameter $\lambda$, 
\begin{equation}
\lambda \equiv \frac{J|E|^{1/2}}{GM^{5/2}},
\end{equation}
where $J$, $E$, and $M$ are the total angular momentum, energy, and mass of the system, and $G$ is the 
gravitational constant.  The (specific) angular momentum per unit mass is 
\begin{equation}
\mathbf{j} = \frac{1}{N}\sum_{i=1}^{N}r_i\times v_i, 
\end{equation}
with $r_i$ and $v_i$ being the position and velocity of $i$-th particle relative to the halo center and halo 
center of momentum, respectively; $N$ is again the total number of particles inside the halo.  The kinetic, 
$E_k$,  and potential, $E_p$, energies are calculated on-the-fly during the simulation as 
\begin{eqnarray}
E_k &=& \frac{1}{2}\sum_{i=1}^{N}m_iv_i^2, \nonumber\\
E_p &=& \left(\frac{N^2-N}{N^2_u-N_u}\right)\left(\frac{-Gm_p^2}{\eta}\right)\sum_{i=1}^{N_u-1}
\sum_{j=i+1}^{N_u}-W(r_{ij}/\eta),
\end{eqnarray}
where $\eta$ is the softening length and $W(u)$ is the softening kernel. If the halo contains more than 1000 particles, 
the potential is calculated using $N_u=1000$ randomly selected particles (for more details see \citealt{Springel2005, 
Bett2007}). 

%%%%%%%%%%%%%%%%%%%%%%%%%%%%%%
\section{Results} 
\label{sec_results}
\begin{figure}
\psfrag{z = 11}[c][c]{$ z =11 ~$}
   \psfrag{z = 16}[c][c]{$ z =16 ~$}
\psfrag{xlab}[c][c]{$\log \lambda$}
\psfrag{ylab}[c][c]{$\rm P (\log \lambda)$}
     \psfrag{-3.0}[c][c]{$-3.0$}
     \psfrag{-2.5}[c][c]{$-2.5$}
     \psfrag{-2.0}[c][c]{$-2.0$}
     \psfrag{-1.5}[c][c]{$-1.5$}
     \psfrag{-1.0}[c][c]{$-1.0$}
      \psfrag{-0.5}[c][c]{$-0.5$}
               \psfrag{0.0}[c][c]{$0$}
            \psfrag{0.1}[c][c]{$0.1$}
               \psfrag{0.2}[c][c]{$ 0.2 $}
      \psfrag{0.3}[c][c]{$0.3 $}
     \psfrag{0.4}[c][c]{$0.4$}
        \psfrag{0.5}[c][c]{$ 0.5$}
     \psfrag{0.6}[c][c]{$ 0.6 $ }
        \psfrag{0.7}[c][c]{$0.7$}
     \psfrag{0.8}[c][c]{$0.8 $}
 \psfrag{0.9}[c][c]{$0.9 $}
 \psfrag{1.0}[c][c]{$1.0 $}
 \psfrag{1.1}[c][c]{$1.1 $}
 \psfrag{1.2}[c][c]{$1.2 $}
 \psfrag{1.3}[c][c]{$1.3 $}
  \psfrag{1.4}[c][c]{$1.4 $}

\centering
\begin{tabular}{cc}
 \includegraphics[trim = 4mm 4mm 4mm 18mm, clip, width=1\columnwidth]{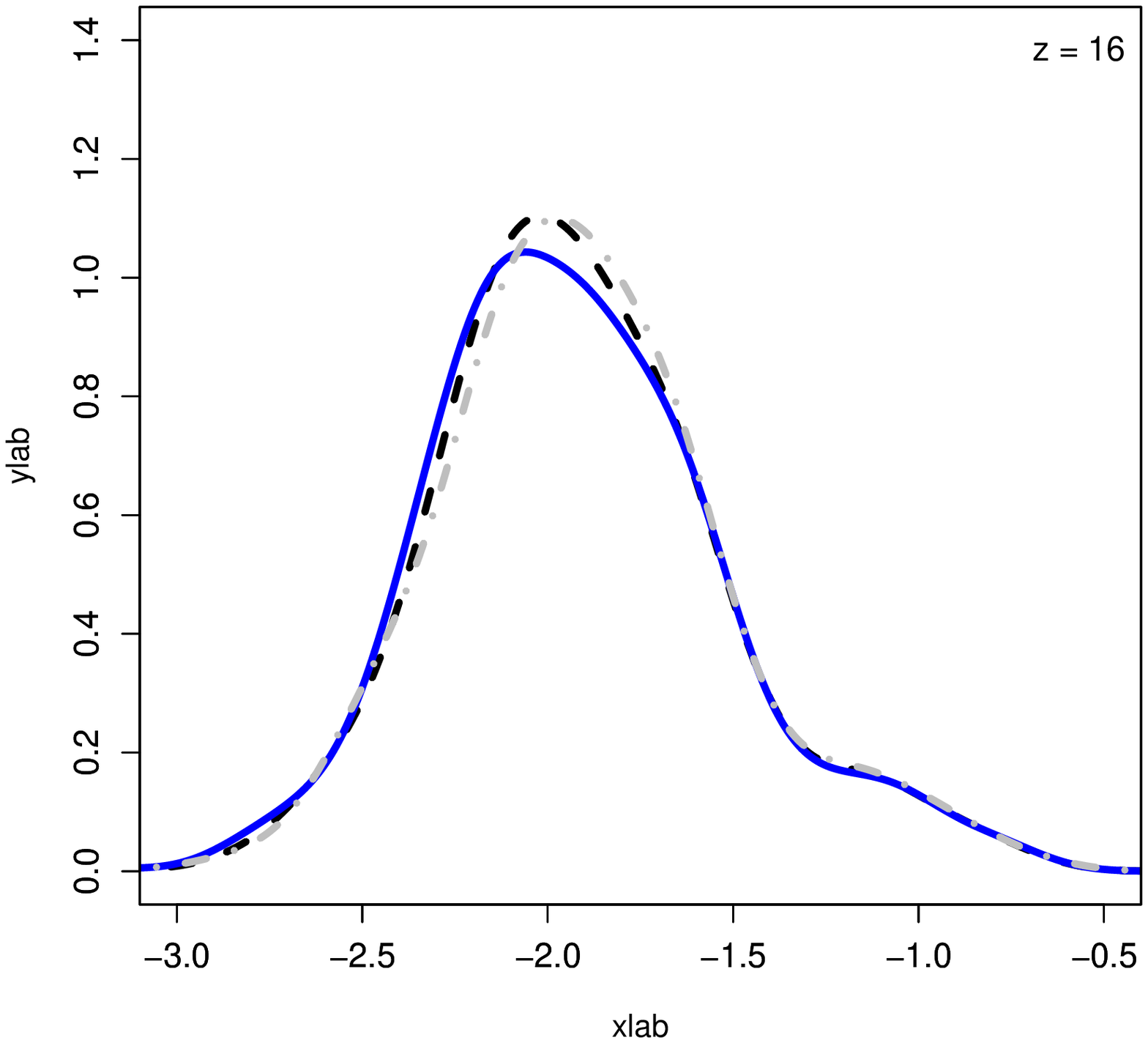}\\
  \includegraphics[trim = 4mm 4mm 4mm 10mm, clip, width=1\columnwidth]{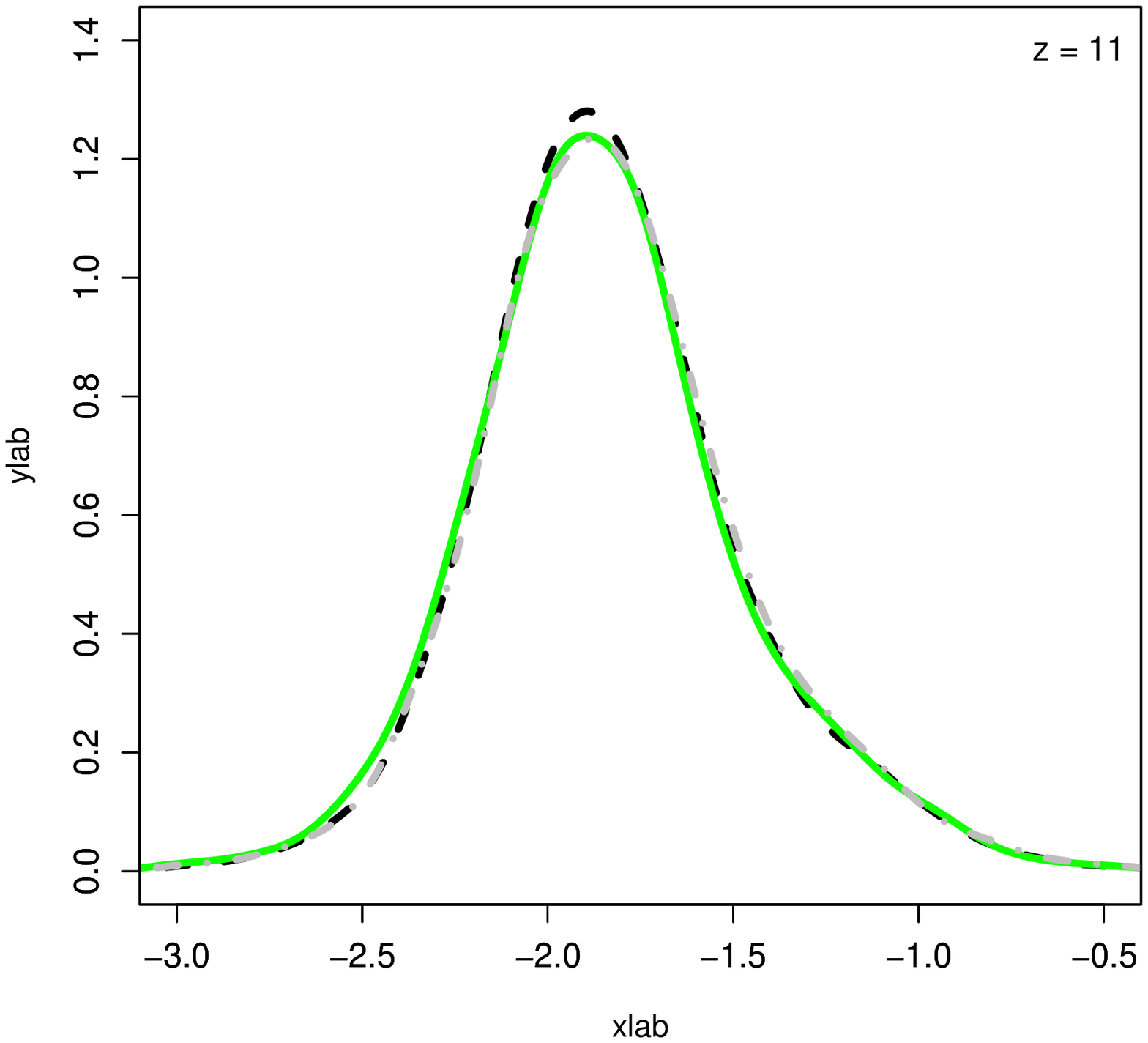}&
\end{tabular}
\caption{Spin distribution of the total (dark matter + gas) halo (black dashed lines), the dark matter
(grey dot-dashed lines) and the gas (solid lines). The upper and lower panels refer to $z=16$ and 11,
respectively.}
\label{fig:spin16}
\end{figure}
\subsection{Shape distribution}

The shape dependence on halo mass has been considered previously by several authors. Despite they all agree on
the non-spherical nature of halos, overall conclusions can be different depending on the assumptions made to define 
halos, the methods to measure shapes or the inclusion of gas physics \citep{Allgood2006}. \citet{Bett2007} found that 
more massive halos tend to be less spherical and more prolate; \citet{Kazantzidis2004} noticed that 
halos formed in simulations with gas cooling are more spherical than halos in adiabatic simulations.

To estimate the halo shape, we only use  halos with more than 100 particles (gas + dark matter), 
equivalent to a total mass of
$\approx 10^{4-5} M_{\odot}$.  The shape of the halos is described in terms of sphericity, $s$, and triaxiality,   
$T$, as defined in Sec. \ref{sec_data}. The probability contour levels of $s$ and $T$ as a function of the total halo mass, 
$M_h$, are shown in Fig. \ref{fig:shape16} for $z = 16$ (upper panel)  and $z=11$ (lower panel).   

In the entire redshift range considered ($11 < z < 16$) the average sphericity  is $\langle s \rangle = 0.3 \pm 0.1$, 
and for more than 90\% of halos  $T \lesssim 0.4$, showing a clear preference for oblateness over prolateness. 
 This is markedly different from
$z=0$ halos that tend to be more prolate and spherical: for example, pure collisionless simulations \citep{Allgood2006}
found $\langle s \rangle \approx 0.6 \pm0.1$ for galaxy mass halos $\sim 10^{12} h^{-1} M_{\odot}$.  Note that constraints from observations of the Sagittarius tidal streams give best-fit parameters 
$s \approx 0.67$, $q \approx 0.83$, and $T \approx 0.56$, in agreement with the Galactic model by
\citet{Law2009}.

Using a non-linear last square method computed with the Gauss-Newton algorithm, and approximating 
$s \propto M_h^{\alpha_s}$ and $T \propto M_h^{\alpha_T}$, we find  
$\alpha_s \approx 0.147$ and  $\alpha_T= 0.285$ at \emph{z} = 16, and $\alpha_s \approx 0.128$ and 
$\alpha_T= 0.276$ at \emph{z} = 11. The fits are given  below\footnote{Note that all fits hereafter are valid only within the redshift range of the simulation, unless explicitly stated. }, 
\begin{eqnarray}
\langle s (z) \rangle &=& \zeta_s \left(\frac{M_h}{h^{-1} M_{\odot}}\right)^{\alpha_s}, \\
\alpha_s &=& -4.1082 +0.9834z-0.0754z^2+0.00191z^3,\nonumber\\
\zeta_s &=& 20.143-4.977z+0.384z^2-0.00979z^3,\nonumber
\label{eq:SM}
\end{eqnarray}
and
\begin{eqnarray}
\langle T(z) \rangle &=& \zeta_T \left(\frac{M_h}{h^{-1} M_{\odot}}\right)^{\alpha_T}, \\
\alpha_T &=& -9.7534 +2.2924z-0.1727z^2+0.00429z^3,\nonumber\\
\zeta_T &=& 51.720-12.500z+0.946z^2-0.0236z^3.\nonumber
\label{eq:TM}
\end{eqnarray}
The most massive halos in the simulation ($M_h=10^{6.5-7} M_{\odot}$) tend to be more spherical and prolate than smaller ones, with 
weak variation of the mass dependence with redshift. 

%%%%%%%%%%%%%%%%%%%%
\subsection{Spin distribution}
\label{sec:spin_dist}
%%%%%%%%%%%%%%%%%%%%

The spin distribution of the  total (dark matter + gas) mass,  dark matter mass and gas mass is shown in Fig. 
\ref{fig:spin16}   for  $z=16$ (top panel) and $z=11$ (low panel). 
The curve is  smoothed using a kernel density estimator  for a sample of n elements, 
\begin{equation}
f(x,h_s) = \frac{1}{nh_s(x)}\sum_{i=1}^{n}K\left(\frac{x-x_i}{h_s(x)}\right), 
\end{equation}
with  a Gaussian kernel $K$ and  an adaptive bandwidth $h_s$.  
The distributions of dark matter and gas are considerably different at high redshift ($z=16$), 
with the baryons rotating slower than the dark matter, which gives the dominant 
contribution to the total spin.  At lower redshift, instead, the spin distributions of dark matter and gas 
track each other almost perfectly, 
as a consequence of a longer time interval available for momentum redistribution between the two components. It is 
important to notice that  the comparisons are done by collecting different  dark matter and gas particles of the same halo,
thus fully accounting for the back-reaction of baryons on the parent dark matter distribution. 
\begin{figure}
\psfrag{lognormal}[c][c]{Lognormal}
\psfrag{Spin}[c][c]{$\log \lambda$}
\psfrag{CDF}[c][c]{CDF}
\psfrag{5e-04}[c][c]{$ $}
\psfrag{1e-03}[c][c]{$ $}
\psfrag{2e-03}[c][c]{$ $}
\psfrag{5e-03}[c][c]{$ $}
\psfrag{1e-02}[c][c]{$ $}
\psfrag{2e-02}[c][c]{$ $}
\psfrag{5e-02}[c][c]{$ $}
\psfrag{1e-01}[c][c]{$ $}
\psfrag{2e-01}[c][c]{$ $}
\psfrag{5e-01}[c][c]{$ $}
\psfrag{-4}[c][c]{$-4$}
\psfrag{-3.5}[c][c]{$-3.5$}
\psfrag{-3}[c][c]{$-3$}
\psfrag{-2.5}[c][c]{$-2.5$}
\psfrag{-2}[c][c]{$-2$}
\psfrag{-1.5}[c][c]{$-1.5$}
\psfrag{-1}[c][c]{$-1$}\
\psfrag{-0.5}[c][c]{$-0.5$}
\psfrag{0.0}[c][c]{$0$}
\psfrag{0.2}[c][c]{$0.2$}
\psfrag{0.4}[c][c]{$0.4$}
\psfrag{0.6}[c][c]{$0.6$}
\psfrag{0.8}[c][c]{$0.8$}
\psfrag{1.0}[c][c]{$1.0$}
    \includegraphics[trim = 4mm 4mm 4mm 18mm, clip, width=1\columnwidth]{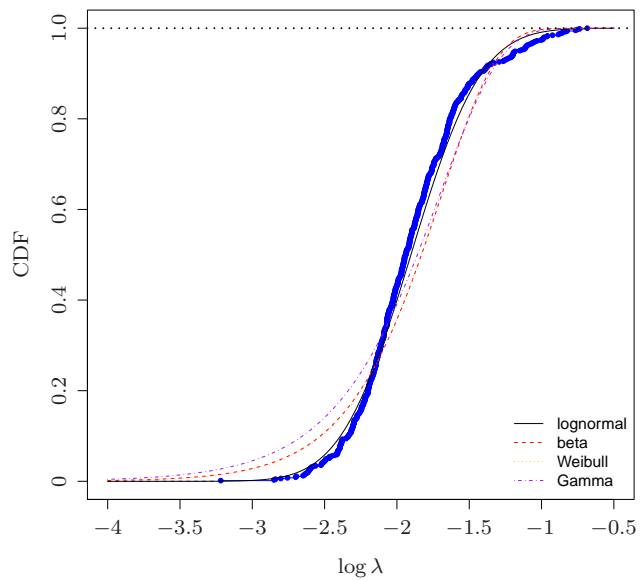}
    \caption{Cumulative distribution function of dark matter spin (thick blue solid line) compared with theoretical distributions at 
$z= 16$.  The theoretical distributions are: lognormal (black solid line), beta (red dashed), Weibull (orange dotted) and Gamma (purple dot-dashed). } 
\label{fig:CDF_11}
\end{figure}
It is very common to fit the halo spin distribution with a lognormal function 
\citep{Bailin2005,Davis2010}.  However, \citet{Bett2007} found deviations from such functional form 
when studying a large ($> 10^6$) number of halos in the   Millennium simulation.
Both the lognormal and Weibull models can be used quite effectively to analyze skewed data sets. 
Although these two models may provide similar data fit for moderate sample sizes, the inferences based on the 
model will often involve tail probabilities, where the effect of the model assumptions is very critical. This  
makes it important to use a quantitative diagnostic to quantify the best distribution to use. 
To do so, we test four classical distributions:  lognormal, Gamma, beta and Weibull, whose shapes are given by:
\begin{eqnarray}
f_{L }(x;\mu,\sigma) &=& \frac{1}{x\sigma\sqrt{2\pi}}e^{-(\ln x -\mu)^2/{2\sigma^2}},\\
f_{\Gamma}(x;k,\theta) &=& \frac{1}{\theta^k}\frac{1}{\Gamma(k)}x^{k-1}e^{(1-x)/\theta},\\ 
f_{\beta}(x;\alpha,\beta) &=& \frac{\Gamma(\alpha+\beta)}{\Gamma(\alpha)\Gamma(\beta)x^{\alpha-1}(1-x)^{\beta-1}},\\
f_{W}(x; k, \lambda) &=& \frac{k}{\lambda}\left(\frac{x}{\lambda}\right)^{k-1}e^{-(x/\lambda)^k}.
\end{eqnarray}
In order to choose the most suitable distribution, we fit the spin distribution of halos in our redshift range. Then, we use   
a Maximum Likelihood test to justify our choice. We obtain the following redshift-averaged values 
of the reduced chi-square: ($\chi^{2}_{L}, \chi^{2}_{\beta}, \chi^{2}_{\Gamma}, \chi^{2}_{W}) =  (2.89 \pm 0.94, 11.95 
\pm 4.85, 11.38 \pm 4.85, 13.53 \pm 4.34$).  In Fig. \ref{fig:CDF_11},  
we plot the cumulative distribution function for the four distributions above and the one obtained from
the simulations. It is clear that the lognormal fits the data at best.  

The distribution of spin parameter can be written as
\begin{equation}
P(\lambda) = \frac{1}{\lambda\sigma_0\sqrt{2\pi}}\exp{\left[-\frac{\ln^2{(\lambda/\lambda_0)}}{2\sigma_0^2}\right]},
\end{equation}
where $\lambda_0$ is the location parameter and $\sigma_0$ is the shape parameter of the distribution. The best fit values 
for mean and variance, $\sigma^2$,  as a function of redshift are\footnote{The mean and variance of 
lognormal distribution are given by $\langle{\lambda}\rangle = e^{\lambda_0+\sigma_0^2/2}$  and $\sigma^2 = e^{\sigma_0^2+2\lambda_0}(e^{\sigma^2_0}-1)$.} 
\begin{eqnarray}
\langle{\lambda}(z) \rangle &=& 1.315-3.681\times10^{-1}z+3.903\times10^{-2}z^2\nonumber \\
&-&1.831\times10^{-3}z^3+ 3.206\times10^{-5}z^4,\\
\rm and\nonumber\\
{\sigma^2}(z) &=&0.1754-5.246\times 10^{-2}z+5.871\times10^{-3}z^2\nonumber \\
&-&2.909\times10^{-4}z^3+5.385\times10^{-6}z^4.
\end{eqnarray}

\begin{figure}
\psfrag{xlabel}[c][c]{$\log{M_{h}} ~(M_{\odot})$}
\psfrag{ylab}[c][c]{$\log \lambda$}
\psfrag{z = 16}[c][c]{$ z = 16~$}
\psfrag{z = 11}[c][c]{$z = 11 $}
\psfrag{M}[c][c]{$\log{M_{h}} ~(M_{\odot})$}
\psfrag{2}[c][c]{$(4.5,5]$}
\psfrag{3}[c][c]{$(5,5.5]$}
\psfrag{4}[c][c]{$(5.5,6]$}
\psfrag{5}[c][c]{$(6,6.5]$}
\psfrag{6}[c][c]{$(6.5,7]$}
\psfrag{7}[c][c]{$(7,7.5]$}
\psfrag{8}[c][c]{$(7.5,8]$}
\psfrag{-3.0}[c][c]{$-3.0$}
\psfrag{-2.5}[c][c]{$$}
\psfrag{-2.0}[c][c]{$-2.0$}
\psfrag{-1.5}[c][c]{$$}
\psfrag{-1.0}[c][c]{$-1.0$}
\psfrag{0.0}[c][c]{$0$}

\centering
\begin{tabular}{cc}
\includegraphics[trim = 5mm 0mm 4mm 18mm, clip,width=1.1\columnwidth]{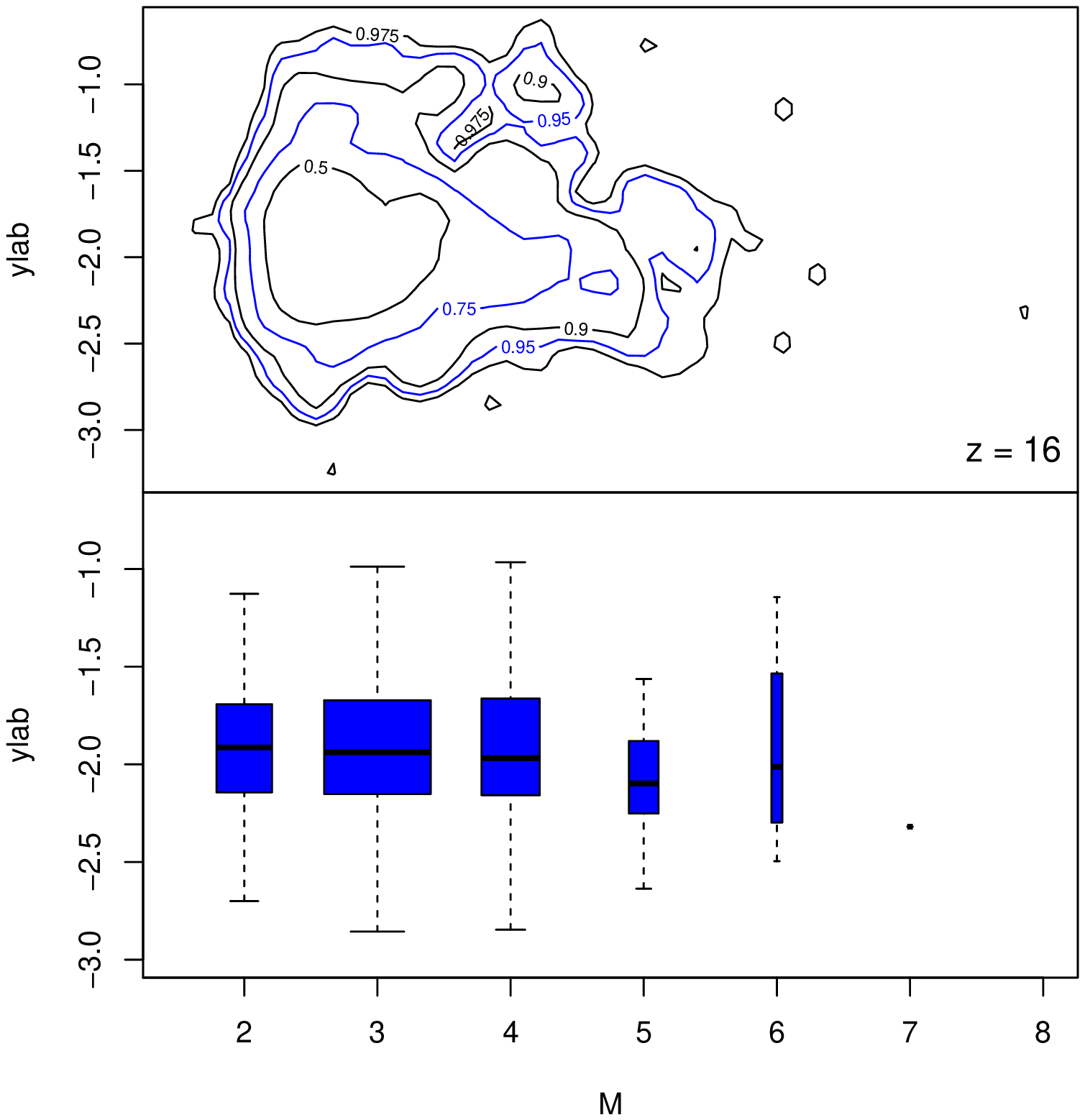}\\
\includegraphics[trim = 5mm 0mm 4mm 10mm, clip,width=1.1\columnwidth]{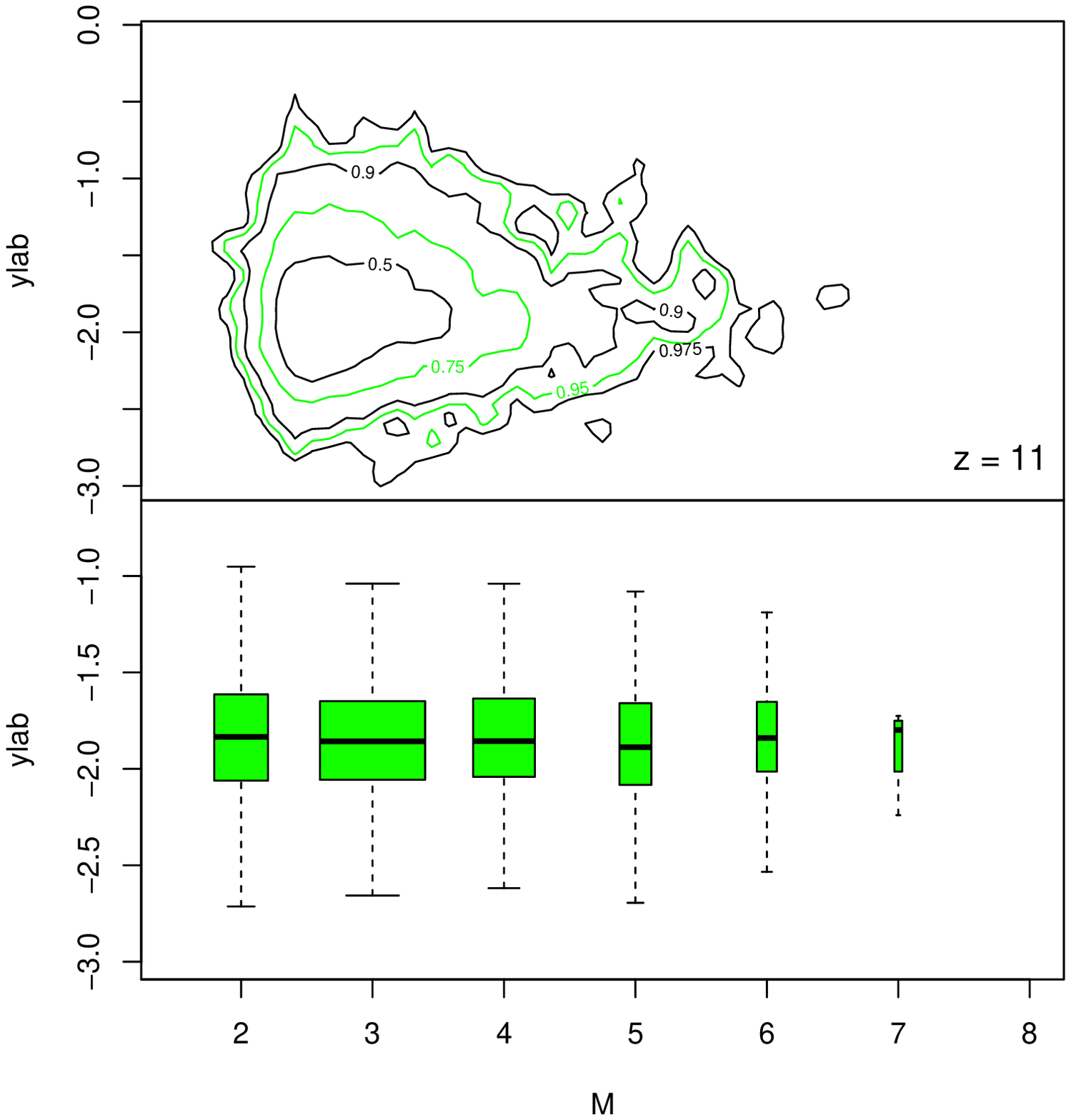} 
\end{tabular}
\caption{Spin distribution as a function of the total mass for halos at $z=16$ (top figure) and 11 (bottom). 
The contour levels in the upper panels represent 50\%, 75\%, 90\%, 95\% and 97.5\% of the sample, 
while in the lower panels 
the median values per bin of mass ($\log{M_h} = 0.5$) is shown. The width of boxes is proportional to the square root of the number 
of halos within each bin and the whiskers extend to the most extreme data point, which is within the 
50\% interquartile range, i.e. the difference between the largest and smallest values in the middle 50\% 
of the dataset.}
\label{fig:spinmass}
\end{figure}

%\begin{eqnarray}
%\langle{\lambda}(z) \rangle &=& 1.315-3.681\times10^{-1}z+3.903\times10^{-2}z^2\nonumber \\
%&-&1.831\times10^{-3}z^3+ 3.206\times10^{-5}z^4,\\
%\rm and\nonumber\\
%{\sigma}(z) &=&4.706-1.405z+1.57\times10^{-1}z^2\nonumber \\
%&-&7.793\times10^{-3}z^3+1.443\times10^{-4}z^4.
%\end{eqnarray}
%If  we include the  a priori  value of $\lambda = 0.0367\pm 0.0429$ \citep{Bett2007} at z = 0,  
%
%\begin{eqnarray}
%\langle{\lambda}(z)\rangle &=& 0.0367-1.330\times10^{-3}z-7.015\times10^{-5}z^2\nonumber \\
%&+&5.166\times10^{-6}z^3,\\
%\rm and\nonumber\\
%{\sigma}(z) &=&0.0429-4.819\times10^{-4}z-3.381\times10^{-4}z^2\nonumber \\
%&+&1.733\times10^{-5}z^3.
%\end{eqnarray}
While it is well known  that the dependence of spin on halo masses is relatively weak \citep{Maccio2007,Bett2007}, it has not been verified yet whether this holds also for small objects. Therefore, analogously to the shape distribution, in Fig. \ref{fig:spinmass} we show the contour levels of the halo 
spin as a function of total halo mass at  $z=16$ and 11. 

Assuming a power law $\lambda \propto M_h^{\alpha}$, best fit becomes 
\begin{eqnarray}
\langle \lambda(z) \rangle &=& \zeta_{\lambda} \left(\frac{M_h}{h^{-1} M_{\odot}}\right)^{\alpha_{\lambda}}, \\
\alpha_{\lambda} &=& -1.64\times10^{2} +4.97\times10^1z-5.60z^2\nonumber\\
&+&2.78\times10^{-1}z^3-5.16\times10^{-3}z^4,\nonumber\\
\zeta_{\lambda} &=& 4.056-0.993z+0.045z^2-0.004z^3.\nonumber
\label{eq:Tmass}
\end{eqnarray}
The slope $\alpha_{\lambda}$ evolves from $-0.023$ at $z=16$, to $0.012$ at $z=11$; both values are consistent
with 0, indicating that the spin parameter distribution is essentially independent on halo mass also at the 
very high redshifts considered here.

%%%%%%%%%%%%%%%%%%%%%%%%%%%%%%%%%%%%%%%%%%
\section{Implications for the PopIII IMF}
\label{sec_IMF}
%%%%%%%%%%%%%%%%%%%%%%%%%%%%%%%%%%%%%%%%%%

In the following we discuss the possibility to connect the host halo properties to the  typical mass scale of the collapsed protostar.  It is important to keep in mind that all results presented here rely on the best available semi-analytical model to translate the 
angular momentum distribution into a corresponding stellar IMF. Despite such limitation, our approach provides the best 
way to statistically analyze a large sample of simulated halos, which could not be performed otherwise. For a detailed study of the properties of the rotation and structure of PopIII stars we refers the reader for the works of     \citet{Greif2012,Stacyb2012}. They  resolve four minihalos down to   scales as small as 0.05 $R_{\odot}$. They found  little evidence of correlation between the properties of each host minihalo and the spin of its largest protostar or the total number of protostars formed in the minihalo. However due the low number of halos probed, it's difficult reach a statistical significant conclusion. 

The IMF is usually  defined  as a segmented power-law or a log-normal type mass distribution \citep{Kroupa2001,Chabrier2003}.  Both definitions  are correlated by 
\begin{equation}
dN \propto m_\star^{-\alpha}dm_\star \quad {\rm or} \quad dN \propto m_\star^{\Gamma} d (\log{m_\star}),
\end{equation}
where $\Gamma = - (\alpha-1)$ \citep{Bonnell2007} and $N$ is  the number of stars with masses in the range $m_\star$ to $m_\star+dm_\star$. The Salpeter slope \citep{Salpeter1955}  is  given by $\alpha = 2.35$, 
or $\Gamma = -1.35$.  As discussed in the Introduction, the PopIII IMF is actually unknown, but there are hints that it
could have been biased towards more massive stars compared to that of PopII/I stars. 

One of the important ingredients for the determination of the IMF is the typical  rotation of the gas. Extreme ultraviolet radiation 
from the protostar can ionize infalling neutral gas, creating an {H \sc{ii}} region whose expansion reduces significantly  the accretion 
of gas onto the star. The radiation would also destroy molecules and inhibit star formation in the surroundings, 
by affecting accretion onto more distant protostellar cores (see  e.g. \citealt{Ricotti2002,Ahn2007,Whalen2008,Petkova2012}).  
The expansion is facilitated by protostellar cores with higher rotation, as the density in the polar directions tends to be lower;
hence the final stellar mass is expected to increase in slowly rotating cores where the effects of feedback are quenched. 

In the following we will make a simple estimate for the expected PopIII IMF, based on a combination of results from our simulations 
and the semi-analytic prescription described in  MT08.  In addition to studying the dependence of the final stellar mass on several  
radiative feedback processes such as  photodissociation of H2, Lyman-$\alpha$ radiation pressure, formation and expansion of 
{H \sc{ii}} regions, and disk photo-evaporation, the authors investigate the role of gas rotation, which modifies the density
distribution in the vicinity of the star.  In their fiducial model they assume that accretion  is halted when the disk photo-evaporation 
rate exceeds the accretion rate onto the star-disk system; at that stage  the typical  PopIII stellar mass is  $\sim 137 M_{\odot}$.

Such value depends strongly on the assumed rotation rate of the protostellar core, which in their case was taken to be half of the keplerian velocity
as suggested by AMR simulations focusing on a single parent halo \citep{abel2002,Bromm2002b,OShea2007}. However,
high rotational speeds might not be the rule, as we have seen from our previous analysis.  \citet{hosokawa2011} have found 
qualitatively similar results by means of a 2D hydrodynamic, radiative transfer simulation,  showing  that the mass accretion  is 
shut down due to the dynamical expansion of the {H\sc{ii}} region and the photo-evaporation of the circumstellar disk. However,  
their simulations found systematically lower final masses than  MT08. The likely explanation suggested by  \citet{hosokawa2011} 
for this discrepancy comes from the different model of the  stellar feedback. In addition,  MT08 assume that after the formation of 
the {H\sc{ii}} region mass accretion onto the  disk still continues from regions shaded by the disk, which is not in agreement with 
the simulation picture from \citet{hosokawa2011}.

In the following, we consider  the influence of rotation in two different scenarios developed by MT08. In the first
(hereafter MT1),  the accretion is reduced by the expansion of the {H \sc{ii}} region  around the protostar and sets a typical 
mass similar to the values found by \citet{hosokawa2011}. In the second (hereafter MT2),  the accretion efficiency is reduced as the {H \sc{ii}}  region expands, however, accretion is allowed to continue from directions that are shadowed by the disk photosphere. The  accretion stops when  the 
photo-evaporation rate exceeds the accretion rate onto the star-disk system and sets a larger typical mass scale. The comparison 
between our results  and MT08 can be easily done due to the one-to-one relationship  between $f_{kep}$ 
defined below and the stellar mass scale, 
assuming other parameters fixed (see Fig. 5, 10 and Table 1 in MT08, where the values to interpolate the one-to-one relationship were taken).

The angular momentum of the gas accreting onto the star-disk system can be  characterized by its keplerian parameter:
\begin{equation}
f_{kep}(M_{r_i}) = V_{\phi,{i}}/V_{kep,{i}},
\end{equation}
where $V_{\phi,{i}}$ and  $V_{kep,{i}}$ are the rotational and keplerian velocity, respectively, as a function of
the total mass enclosed within a radius $r_{i}$. Here,  $r_i$ is the position vector of the $i$-th particle 
relative to the halo centre. The characteristic velocities above are defined as:
\begin{equation}
V_{\phi,i} = j_{i}/r_i, \quad V_{kep,i} = \sqrt{GM(r_i)/r_i},
\end{equation}
where  the specific angular momentum, $j_{i}$, of the $i$-th  particle is  averaged over the spherical shell whose radius is $r_i$. 

The value of $f_{kep} $ averaged over all  particles within a halo can be expressed as
\begin{equation}
\langle f_{kep} \rangle=\frac{1}{N}\sum_{i=1}^{N} \frac{(j_{i}/r_i)}{V_{kep,i}}.
\label{eq:avg_fkep}
\end{equation}
According to \cite{mckee2008},  for $f_{kep} \gtrsim 0.25$ little difference is observed in the final stellar  
mass, which is set by the balance between the (inner) disk-shadowed accretion and mass loss due to 
photo-evaporation. For smaller rotation parameters ($f_{kep} \lesssim 0.125$) instead, the mass scale
 at which accretion is halted strongly depends on the {H \sc{ii}} region breakout. 

\begin{figure}
\psfrag{z = 16}[c][c]{$z = 16 $}
\psfrag{z = 11}[c][c]{$z = 11 $}
\psfrag{-3.5}[c][c]{$-3.5$}
\psfrag{-3.0}[c][c]{$-3$}
\psfrag{-2.5}[c][c]{$-2.5$}
\psfrag{-2.0}[c][c]{$-2$}
\psfrag{-1.5}[c][c]{$-1.5$}
\psfrag{-1.0}[c][c]{$-1$}\
\psfrag{-0.5}[c][c]{$-0.5$}
\psfrag{0.0}[c][c]{$0$}
\psfrag{0.5}[c][c]{$0.5$}
\psfrag{0.2}[c][c]{$0.2$}
\psfrag{0.4}[c][c]{$0.4$}
\psfrag{0.6}[c][c]{$0.6$}
\psfrag{0.8}[c][c]{$0.8$}
\psfrag{1.0}[c][c]{$1.0$}
\psfrag{xlabel}[c][c]{$\log{\langle f_{kep} \rangle}$}
\psfrag{ylabel}[c][c]{P$(\log{\langle f_{kep} \rangle})$}
\includegraphics[trim = 5mm 0mm 4mm 18mm, clip,width=1\columnwidth]{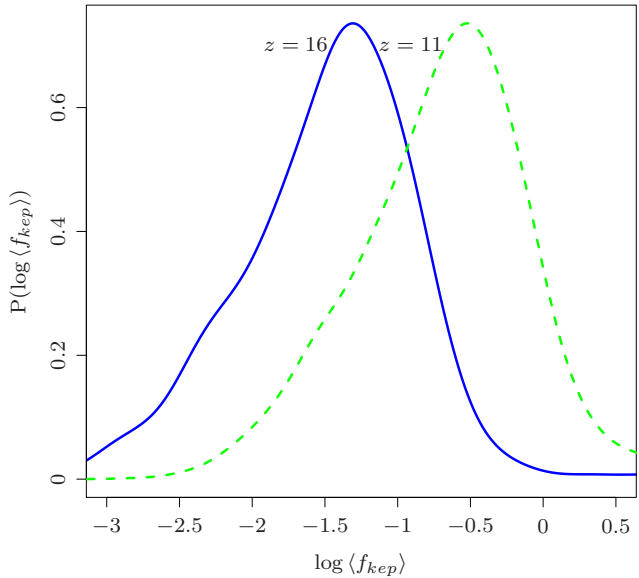}
\caption{Distribution of $\log \langle f_{kep} \rangle$  at  $z = $16 (solid blue line) and 11 (dashed green).}
\label{fig:fitkep}
\end{figure}

To calculate the  $\langle f_{kep} \rangle$  distribution, we use the same procedure described in Sec. \ref{sec:spin_dist} and we show it in Fig. \ref{fig:fitkep} at $z = 16$ and 11.    
The overall  shape of the distribution  remains  
qualitatively similar at different epochs, however the mean  value increases towards lower redshifts, from 0.26 at \emph{z} = 16 to 0.61 at \emph{z} = 11. 

\begin{figure}
\psfrag{ylabel}[c][c]{$\rm P(m_{*}$)}
\psfrag{Msun}[c][c]{$m_{*} (M_{\odot})$}
\psfrag{40}[c][c]{$40$}
\psfrag{60}[c][c]{$60$}
\psfrag{80}[c][c]{$80$}
\psfrag{100}[c][c]{$100$}
\psfrag{120}[c][c]{$120$}
\psfrag{130}[c][c]{$130$}
\psfrag{140}[c][c]{$140$}
\psfrag{150}[c][c]{$150$}
\psfrag{160}[c][c]{$160$}
\psfrag{170}[c][c]{$170$}
\psfrag{0.00}[c][c]{$0$}
\psfrag{0.01}[c][c]{$0.01$}
\psfrag{0.02}[c][c]{$0.02$}
\psfrag{0.03}[c][c]{$0.03$}
\psfrag{0.04}[c][c]{$0.04$}
\psfrag{0.05}[c][c]{$0.05$}
\psfrag{0.06}[c][c]{$0.06$}
\psfrag{0.08}[c][c]{$0.08$}
\centering
\begin{tabular}{cc}
\includegraphics[trim = 5mm 0mm 4mm 18mm, clip,width=1\columnwidth]{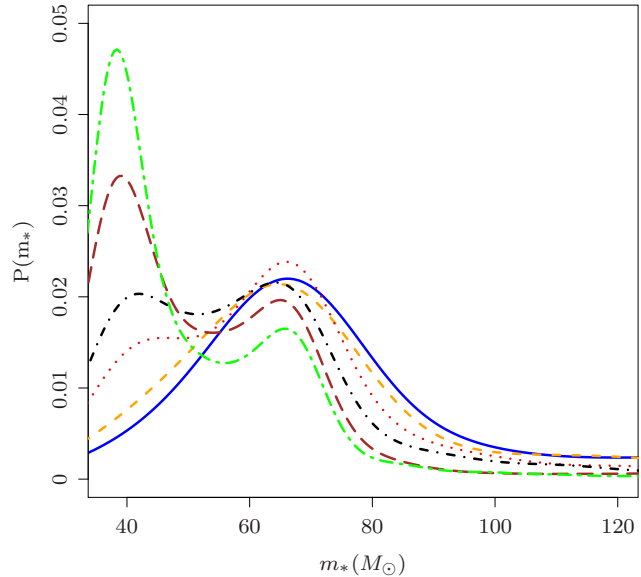}\\
\includegraphics[trim = 5mm 0mm 4mm 14mm, clip,width=1\columnwidth]{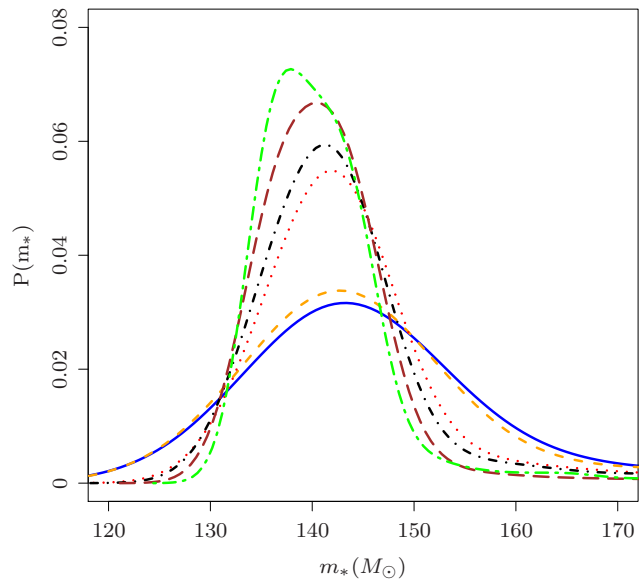}
\end{tabular}
\caption{PopIII typical mass distribution for model MT1 (top panel)  and MT2 (bottom panel) at redshift  \emph{z} =16 (blue solid line), 15 (orange dashed), 14 (red dotted), 13 (black dashed-dot), 12 (brown long-dashed), and 11 (green long dashed-dot). } 
\label{fig:IMFHII}
\end{figure}

The typical rotational velocity is  in general  below the required velocity for rotational support, in agreement with 
previous calculations \citep{abel2002}. Using the above distributions we can translate the rotational
velocity into a PopIII typical mass using models MT1 and MT2. The result of this exercise is shown in 
Fig. \ref{fig:IMFHII}; there are several interesting features that we can deduce from here.

First, at high redshift the IMF tends to closely track the lognormal distribution imprinted by the rotation
properties of the halos. Depending on the feedback model, though, the distribution can be centered at  $\approx 65 M_\odot$ (MT1)
or $\approx 140 M_\odot$ (MT2). At later times, model MT1 tends to evolve into a bimodal distribution 
with a second prominent peak located at $35-40 M_\odot$ in addition to the initial one. The bimodality comes from the non-linear connection between rotation and mass scale. 
For values of $f_{kep} \gtrsim 0.25$ the rotation has a weak influence on the final stellar mass.  
%{\bf BC: to which model does the following sentence refer to? it's not clear, because if it's MT1 there
%are two peaks and the second doesn't really look to be shrinking}
As the redshift decreases, so does the width around the  second  peak of the stellar mass distribution, because the majority of halos 
have higher values of $f_{kep}$. A peak at $m_\star \sim 65 M_\odot$ is still present due to the slow rotation
tail of the $\langle f_{kep} \rangle$ distribution.
Model MT2 instead shows a much more gradual and moderate shift of the peak towards lower masses, accompanied by 
an increasingly narrower width. 

Thus, it seems that the $\langle f_{kep} \rangle$ distribution and shift with redshift,
governed by the angular momentum evolution of the halos, has an 
extremely strong influence on the PopIII IMF, at least as long as we assume the MT08 feedback models
to be correct.
However, it is important to emphasize that the gas rotation is also affected by thermal heating from feedback
mechanisms, which are expected to be stronger for PopIII stars and in low mass halos, as the ones we
are dealing with. In addition, simulations (e.g. \citealt{tornatore2007,maio2010,Maio2011}) show that it is common to
have multiple star formation sites within the same halo, with a combination of PopIII and PopII/I stars. This
means that it is not straightforward nor trivial to assign a single PopIII IMF to a halo.

With the above caveats in mind, we come to the somewhat surprising conclusion that, although on a protostellar basis radiative feedback acting
on baryons might be the key factor in determining the mass of the first stars, \emph{it is the angular momentum distribution 
of the dark matter halos that controls the build-up of the IMF} (see also \citealt{Schroyen2011}). 
This process might work in the simple way outlined here 
as long as there is a one-to-one correlation between the halo and the protostellar core angular momentum and it might
break down in larger galaxies in which momentum is dissipated via tidal torques and/or shocks arising from the interaction
among different cores or galactic-scale dynamical instabilities. We reiterate that all our calculations include the back-reaction
of baryons on the dark matter, and hence they should provide a robust description of the total matter dynamics in a halo.

%%%%%%%%%%%%%%%%%%%%%%%%%%%%%%%%%%%%%%%%%%%%%%%%%%%%%%%%%%%%%%%%%%%%%%%%%%%%%%%%%%%%%%%%
\section{Summary}
\label{sec_disc}

Our study, following the evolution of dark matter \emph{and} baryonic physics in cosmological simulations,
makes possible to study the statistical properties of the high-$z$, low mass halos that likely hosted the first stars.
In addition such simulations include a large number of physical processes (PopIII and PopII/I star formation, metal 
enrichment, gas cooling from resonant and fine-structure lines  and feedback effects) and a detailed chemical
network following the abundances of key species (e$^-$, H, H$^+$, H$^-$, He, He$^{+}$, He$^{++}$, H$_2$, 
H$_2^+$, D, D$^+$, HD, HeH$^+$).  

In this work we have mostly concentrated on the statistical analysis of two important halo properties, i.e. spin and shape. 
As these parameters, governing the overall evolution of protostellar cloud collapse, are predicted by modern PopIII formation 
theories to be related to the mass of the first stars forming in these systems, we then discuss the implications of
our findings for the PopIII IMF. Our main results can be summarized as follows:

\begin{itemize}

\item In the entire redshift range considered ($11 < z < 16$) the average sphericity  is $\langle s \rangle = 0.3 \pm 0.1$, 
and for more than 90\% of halos  $T \lesssim 0.4$, showing a clear preference for oblateness over prolateness, contrary 
to what found at $z=0$.

\item Larger halos in the simulation tend to be both more spherical and prolate: we find $s \propto M_h^{\alpha_s}$ and 
$T \propto M_h^{\alpha_T}$, with $\alpha_s \approx 0.128$ and  $\alpha_T= 0.276$ at \emph{z} = 11. 

\item The spin distributions of dark matter and gas are considerably different at $z=16$, with the baryons rotating 
slower than the dark matter (giving the dominant contribution to the total spin).  At lower redshift, instead, the spin 
distributions of dark matter and gas track each other almost perfectly, as a consequence of a longer time interval available 
for momentum redistribution between the two components.

\item The spin distribution for both gas and dark matter inside the simulated small halos can be well represented by a 
lognormal function, with mean and variance at $z=16$ of 0.0184 and 0.000391, virtually independent on halo mass and 
in good qualitative agreement with previous results. The mean value of spin parameters is also in agreement within 1-$\sigma$ with the median value found by  \citet{Jang-Condell2001} for a lognormal distribution in their study of small-scale structures at high-$z$   using  a pure dark matter N-body simulation.  
% \textbf{These are the values I got from eqs. 19-20. How comes that 
%the variance is larger than the mean? Please check (also in the abstract)!}

\item According to most recent theories of PopIII star formation, rotation is the key factor in determining their final mass.
Using the results of two feedback models (MT1 and MT2) by McKee \& Tan (2008) and mapping our halo spin distribution 
into a PopIII IMF, we find that at high $z$ the IMF tends to closely track the spin lognormal distribution; depending on the
feedback model, though, the distribution can be centered at  $\approx 65 M_\odot$ (MT1) or $\approx 140 M_\odot$ (MT2). 
At later times, model MT1 tends to evolve into a bimodal distribution with a second prominent peak located 
at $35-40 M_\odot$, as a result of the non-linear relation between rotation and halo mass.
\end{itemize}

While the PopIII IMF is still highly debated \citep{Stacy2010b,clark2011, greif2011, Prieto2011, hosokawa2011,Greif2012}, the present study
offers an intriguing indication that the IMF of the first stars might be tied and controlled by the properties of
their parent halos, thus linking in a novel way large scale structure and early star formation. If this is indeed the case,
our suggestion could lead to clear and testable predictions (e.g. PISN rates, abundance of pure PopIII
galaxies, metal abundance patterns in the IGM and low-mass stars to mention a few) for the number, properties 
and cosmic evolution of these pristine stellar systems.    

%%%%%%%%%%%%%%%%%%%%%%%%%%%%%%%%%%%%%%%%%%%%%%%%%%%%%%%%%%%%%%%%%%%%%%%%%%%%%%%%%%%%%%
\section*{Acknowledgements}
RSS  and BC  are happy to thank the  Scuola Normale Superiore (SNS), Pisa, Italy  for hospitality during part of the 
development of this work.  RSS acknowledges financial support from the Brazilian financial agency FAPESP through 
grant number 2009/05176-4. UM acknowledges the computing center of the Max Planck Society, in
Garching bei M\"unchen (Rechenzentrum Garching, RZG), Germany, for the
invaluable technical support, and kind hospitality at the Italian
computing center (CINECA), the Scuola Normale Superiore in Pisa, Italy,
the University of Bologna, Italy, and the Universidad Aut\'onoma de Madrid,
Spain. He also acknowledges financial contributions from the Project
``HPC-Europa2'', grant number 228398, with the support of the European
Community,
under the FP7 Research Infrastructure Programme. We also thank Emille Ishida,  Andressa Jendreieck and Jongsoo Kim for fruitful comments.  Finally we thank Volker Bromm for the very kindly and fruitful comments as a  referee of the present paper.

\bsp

\label{lastpage}
\end{document}